\documentclass[twocolumn,secnumarabic,amssymb, nobibnotes, aps, prd]{revtex4-2}
\usepackage[T1]{fontenc} 
\usepackage[]{graphicx}
\usepackage{subfigure}
\usepackage{times}

\usepackage{amsmath}
\usepackage{lipsum}
\usepackage{verbatim}

\usepackage{changes}
\usepackage{bm}
\usepackage{ulem}
\usepackage{xcolor}
\usepackage{lineno}
\usepackage{amsmath}

\begin{document}

\title{\LARGE \bf
Hot-electron-injection-induced symmetry breaking in bilayer MoS$_2$ probed by second-harmonic generation}

\author{Zhizi Guan$^{1,\dagger}$, Zhiwei Peng$^{2,\dagger}$, David J. Srolovitz$^1$, Jacob Khurgin$^3$}

\author{Dangyuan Lei$^{2}$}

\email{dangylei@cityu.edu.hk}

\affiliation{\mbox{$^1$Department of Mechanical Engineering, The University of Hong Kong, Pokfulam Road, Hong Kong SAR}\\\mbox{$^2$Department of Materials Science and Engineering, Centre for Functional Photonics, and Hong Kong} Branch of National Precious Metals Material Engineering Research Centre, City University of Hong Kong, Hong Kong SAR \\$^3$Electrical and Computer Engineering Department, Johns Hopkins University, Baltimore, MD, USA\\}

\begin{abstract}

Symmetry governs the selection rules of light-matter interactions in crystalline materials, making symmetry manipulation a powerful tool for tuning their optical properties. Here, we demonstrate that the hot-electron injection from a plasmonic resonator breaks the centrosymmtry of an adjacent transition metal dichalcogenide bilayer, probed via second-harmonic generation (SHG) in a Au-nanoparticle@bilayer-MoS$_2$@Au-film hybrid system. Power-dependent SHG measurements exhibit saturation behavior, consistent with a capacitor model where interfacial charge accumulation creates a dynamic barrier limiting further electron injection. Polarization-resolved SHG measurements reveal anisotropic second-order susceptibility response under hot-electron injection, where the contrast between different susceptibility components provides a quantitative measure of symmetry-breaking anisotropy. First-principles calculations elucidate the nonlinear optical responses evolution in bilayer MoS$_2$ and comfirm the anisotropic modification of  susceptibility components under hot-electron injection, modeled by a perpendicular electric field. Our work establishes SHG as an effective probe of hot-electron-induced symmetry breaking in 2D materials, demonstrating a promising approach for ultrafast manipulation of material properties through controlled charge injection at the nanoscale.

\end{abstract}

\maketitle


 

\begin{figure*}[htb]
    \centering
    \hspace{-0.23cm}\includegraphics[scale=0.48]{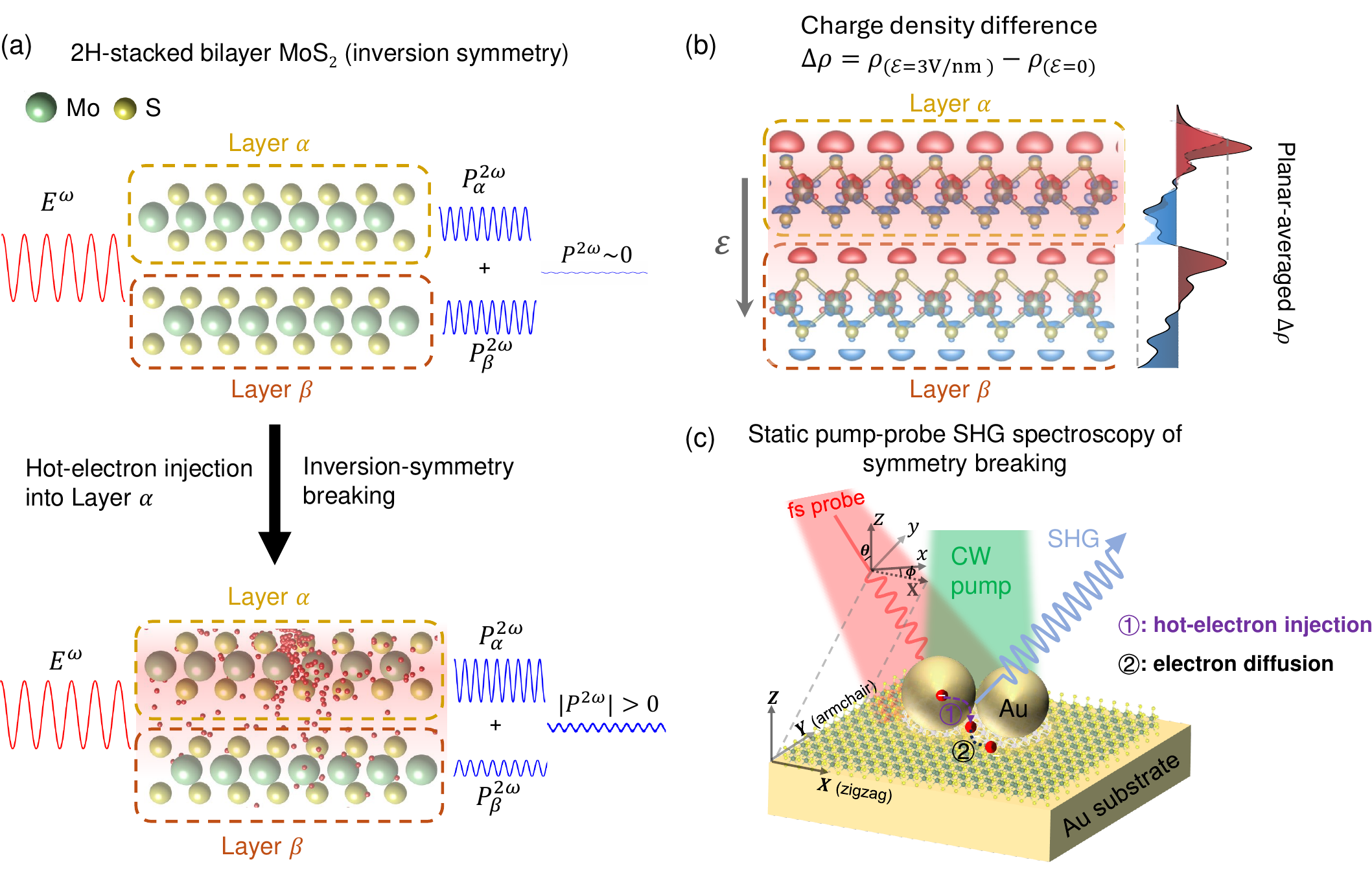}
    \caption{(a) Schematic illustration of SHG probing inversion symmetry breaking in bilayer MoS$_2$ under hot-electron injection onto the upper layer (layer $\alpha$). In pristine 2H-stacked bilayer MoS$_2$ (top), the second-order polarization $P^{2\omega}$ vanishes due to the cancellation between two layers with opposite orientations. Hot-electron injection (bottom) breaks the inversion symmetry, generating non-zero second-order polarization. (b) DFT-calculated charge density difference ($\Delta\rho = \rho_{(\mathcal{E} = 3 \text{ V/nm})} - \rho_{(\mathcal{E} = 0)}$) showing electron accumulation (red) and depletion (blue) regions under an external electric field (strength $\mathcal{E}$), with planar-averaged values in the inset. The asymmetric charge distribution between layers indicates symmetry transformation from $D_\text{3d}$ to $C_\text{3v}$, analogous to hot-electron-induced effects in the bottom panel of Fig.~\ref{Figure1}(a). (c) Experimental configuration of static pump-probe SHG spectroscopy using continuous-wave laser to generate hot electrons and femtosecond laser to induce SHG in the AuNS-dimer@bilayer-MoS$_2$@Au-film heterostructure. Crystal coordinate system ($X$-$Y$-$Z$), laboratory coordinate system ($x$-$y$-$z$), incident polarization angle $\theta$, and azimuthal angle $\phi$ are shown.}
    \label{Figure1}
\end{figure*}

Symmetry plays a key role in the fundamental interactions between light and matter (atoms, molecules,  solids), determining selection rules, optical activity, and nonlinear responses \cite{yu2015charge, klein2017electric, PhysRevB.101.241106, PhysRevLett.127.237402, du2021engineering}. 
Two-dimensional (2D) transition metal dichalcogenides (TMDCs), in their naturally occurring 2H-stacked form, provide an ideal platform for investigating symmetry-dependent phenomena \cite{mak2010atomically,wang2012electronics, chhowalla2013chemistry}. 
When mechanically exfoliated, these van der Waals materials exhibit layer-dependent symmetry: most even-layered structures possess inversion symmetry while odd-layered ones do not \cite{li2013probing, PhysRevB.87.201401}. 
This intrinsic property makes TMDCs particularly suitable for studying symmetry modifications. 
Among the various approaches employed to modify symmetry in centrosymmetric TMDCs, strain \cite{koskinen2014density, mennel2018optical, li2022symmetry,peng2020strain} and electric fields \cite{klein2017electric, yu2015charge,wu2013electrical, PhysRevB.109.075417} have emerged as particularly  effective methods for modifying optical and electrical properties. 
Strain  can induce nonzero second-order optical susceptibility ($\chi^{(2)}$) components in otherwise centrosymmetric 2D crystals. 
Strain and nonlinear susceptibility are connected via the photoelastic tensor;  second harmonic generation (SHG) can serve as a sensitive probe of the full strain tensor in 2D materials \cite{lyubchanskii2000second,jeong2000strain}. 
In our earlier work \cite{PhysRevB.109.075417}, we reported on the application of electric fields to generate novel out-of-plane $\chi^{(2)}$ tensor components in monolayer TMDCs. 
Electric field-induced second-harmonic generation (EFISH) has been employed in various applications in different materials \cite{bavli1991relationship, lee2020electrically, widhalm2022electric}.

However, both strain and electric field approaches face significant limitations. 
Strain engineering requires complex fabrication techniques and mainly produces static (spatially-fixed) strain fields. 
Electrostatic gating approaches demand high voltages and currents become unsustainable at high frequencies when applied to ultrathin 2D materials. 
For wide-band applications requiring picosecond-scale symmetry control, all-optical methods are more promising than conventional strain or voltage approaches \cite{wen2016large}. 
One such method employs plasmon-induced hot-electron injection to break symmetry in even-layered TMDCs \cite{cai2011electrically,wen2018plasmonic,lei2024new}. 
When surface plasmon polaritons in metal structures decay, they transfer energy to conduction band electrons, generating hot electrons and holes. 
Carriers with sufficient energy to overcome the interfacial barrier get injected into the adjacent 2D material \cite{peng2023plasmonic,khurgin2020fundamental}, creating a non-uniform space charge distribution that is symmetry-breaking. 
This enables second-order optical processes normally forbidden by symmetry. 
While hot-carrier injection significantly modifies 2D material optical properties \cite{wen2018plasmonic,peng2023plasmonic, chen2021bandgap, kang2014plasmonic}, the underlying mechanisms and polarization-dependent characteristics are not yet well-understood. 

In this Letter, we demonstrate symmetry breaking in bilayer MoS$_2$ via localized hot-electron injection, using a SHG technique as a probe. 
Our experimental approach combines continuous-wave (CW) excitations for hot-carrier generation with femtosecond pulses for SHG detection, revealing both the mechanisms and polarization characteristics of the symmetry breaking process. 
Theoretical analysis reveals that the spatially non-uniform distribution of injected carriers creates symmetry-breaking effects analogous to those from an applied electric field. 
Power-dependent measurements show saturation behavior that aligns with a  capacitor model of interfacial charge dynamics. 
Polarization-resolved SHG measurements quantify a $\sim$20\% ratio between specific susceptibility tensor components, establishing this relative strength as a direct metric for symmetry-breaking anisotropy.
First-principles calculations corroborate these experimental findings and elucidate the anisotropic evolution of nonlinear optical responses under hot-electron injection, modeled by a perpendicular electric field. 
These results establish SHG as a sensitive probe of hot-electron-induced symmetry breaking and suggest potential approaches for ultrafast symmetry control in low-dimensional materials.


The physical principles behind hot carrier-induced SHG is shown schematically in Fig.\ref{Figure1}(a). 
The SHG polarization is $P^{2\omega}= \varepsilon_{0}\chi^{(2)}(E^{\omega})^2$, where $\varepsilon_{0}$ is the vacuum permittivity, $E^{\omega}$ is the optical electric field, and $\chi^{(2)}$ is the second-order susceptibility tensor. 
In pristine 2H-stacked bilayer MoS$_2$ with $D_\text{3d}$ symmetry, all $\chi^{(2)}$ components vanish due to the cancellation of second-order polarizations between layers. 
This cancellation arises from the 2H stacking configuration where adjacent layers are rotated by $180^\circ$ relative to one another, resulting in equal but oppositely oriented second-order polarization responses. 
Symmetry breaking occurs when non-uniformly distributed hot carriers accumulate at the metal-TMDC interface, generating an electric field that transforms the material symmetry from $D_\text{3d}$ to $C_\text{3v}$ (Fig.~\ref{Figure1}(b)). 
This transformation enables non-vanishing SHG, analogous to the electric-field-induced symmetry transformation observed in monolayer TMDCs \cite{PhysRevB.109.075417}.

To quantitatively understand this symmetry breaking mechanism, we first analyze the SHG response of bilayer MoS$_2$ under a normal external electric field. 
For fundamental radiation incident at an angle $\theta$ with electric field strength $\mathcal{E}$ onto a sample with $C_\text{3v}$ symmetry, the second-order polarization components $P^{2\omega}$ are (see Supplemental Material \cite{SM} for details):

\begin{equation}\label{eq1}
    \left ( \begin{matrix}
    P_x^{2\omega} \\
    P_y^{2\omega} \\
    P_z^{2\omega}
    \end{matrix} \right ) \propto
    \left ( \begin{matrix}
    d_{15}(\mathcal{E})\sin2\theta - d_{22}\sin3\phi \\
    -d_{22}\cos^2\theta\cos3\phi \\
    d_{31}(\mathcal{E})\cos^2\theta + d_{33}(\mathcal{E})\sin^2\theta
    \end{matrix} \right ),
\end{equation}
where $x-y$ denotes the in-plane directions and $z$ is the out-of-plane direction in the laboratory coordinate frame (Fig.~\ref{Figure1}(c)). 
The coefficients $d_{il}$ are expressed in a contracted notation for $\chi^{(2)}$ tensor components \cite{boyd2008nonlinear}, classified by second-harmonic polarization direction. The in-plane components include $d_{22} (\chi^{(2)}_{yyy}$) and $d_{15} (\chi^{(2)}_{xzx})$, while $d_{31} (\chi^{(2)}_{zxx}$) and $d_{33} (\chi^{(2)}_{zzz}$) represent out-of-plane components. 
$\phi$ is the azimuthal angle between the fundamental radiation polarization and crystal axis ($X-Y-Z$) (details of the coordinate system are as per \cite{PhysRevB.109.075417}).

\begin{figure}[tb]
    \centering
    \hspace{-0.45cm}\includegraphics[scale=0.35]{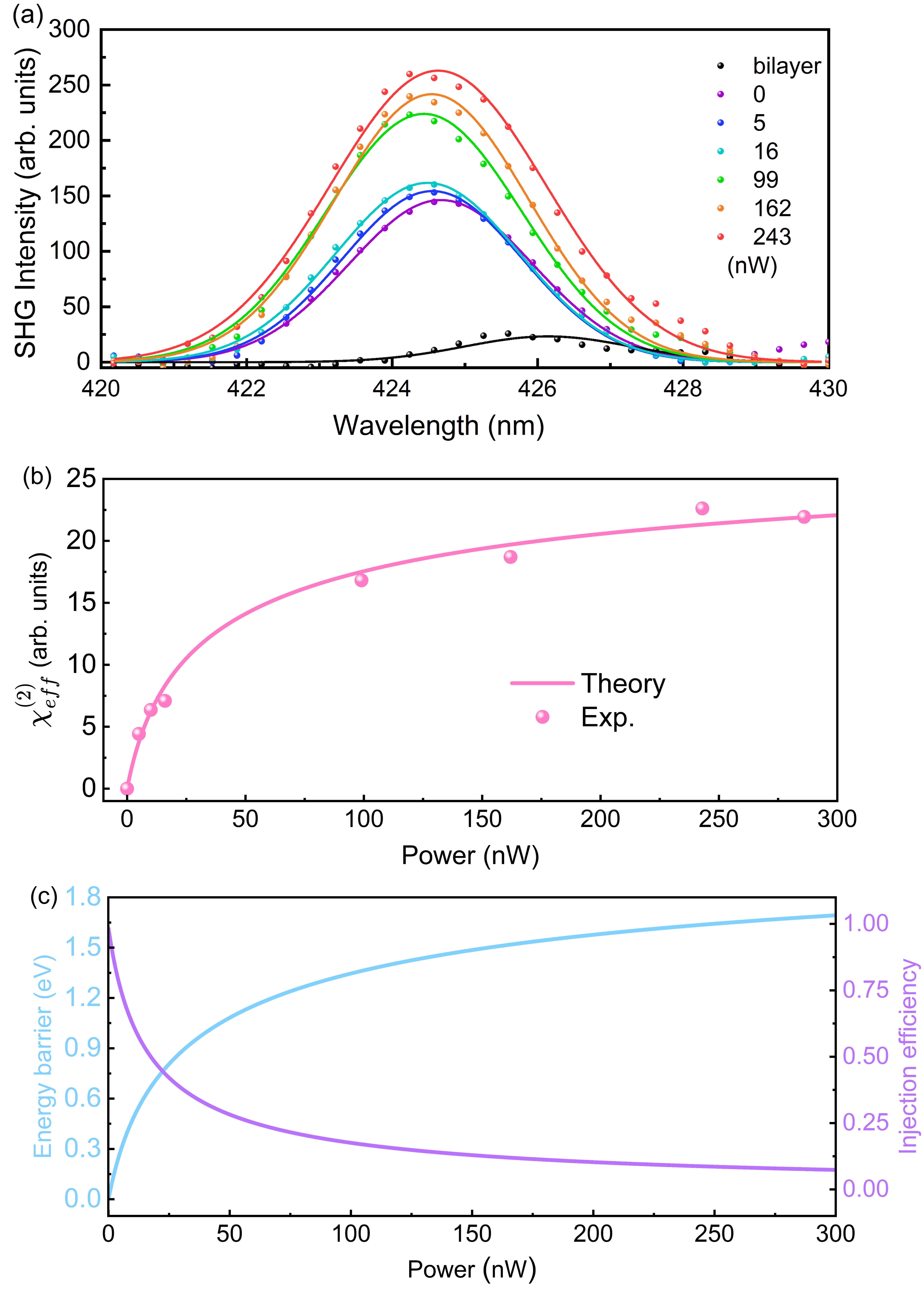}
    \caption{(a) Power-dependent SHG intensity from the AuNS-dimer on a bilayer-MoS$_2$ atop a Au-film structure under 532-nm CW laser excitation. 
(b) Effective second-order susceptibility $\chi^{(2)}_\text{eff}$ extracted from SHG measurements (points) with the capacitor model fit (line). 
(c) Calculated interfacial energy barrier and hot-electron injection efficiency as functions of incident power density.}
    \label{Figure2}
\end{figure}

\begin{figure*}[htb]
    \centering
    \subfigure{\includegraphics[scale=0.68]{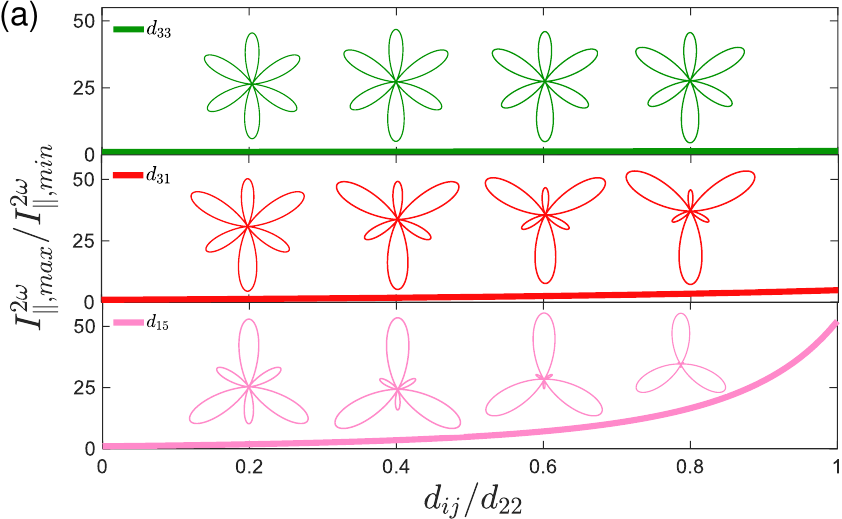}}
    \hspace{1cm}
    \raisebox{0.45cm}{\subfigure{\includegraphics[scale=0.36]{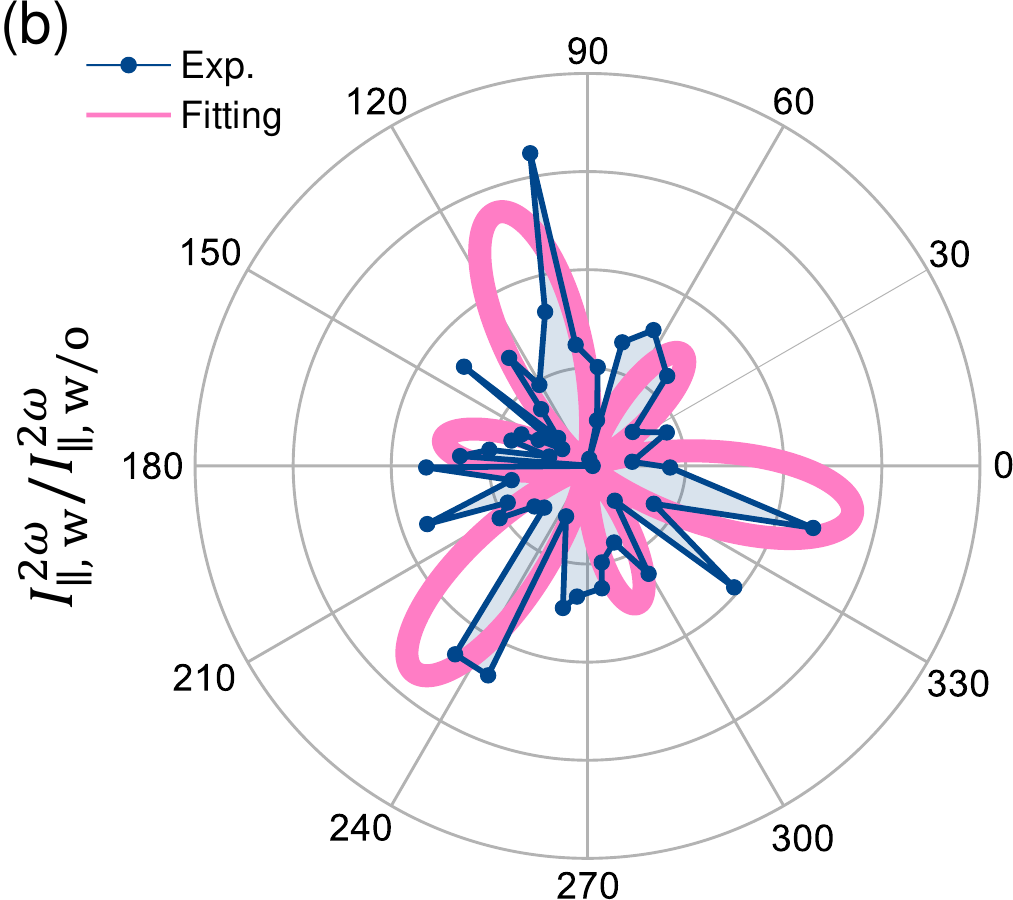}}}
    \caption{(a) Dependence of $I_{\parallel, max}^{2\omega}$/$I_{\parallel, min}^{2\omega}$ on the ratios $d_{ij}/d_{22}$ for different $d_{ij}$ components ($d_{33}$, $d_{31}$, and $d_{15}$). 
For $d_{33}$ and $d_{31}$, maxima occur at $\phi = 30^\circ$ and minima at $\phi = 90^\circ$, while $d_{15}$ shows opposite behavior. 
Polar plots show $I_{\parallel}^{2\omega}$ patterns for $d_{ij}/d_{22}$ = 0.2, 0.4, 0.6, and 0.8. 
(b) Azimuthal dependence of the SHG intensity ratio $I_{\parallel,w}^{2\omega}/I_{\parallel,w/o}^{2\omega}$ from the AuNS-dimer on bilayer-MoS$_2$ atop a Au-film structure with and without 532-nm CW laser excitation which is used to isolate the hot-electron-induced symmetry breaking from the plasmonic background.}
    \label{Figure3}
\end{figure*}

From Eq.~\ref{eq1} and Fig.~S1, the second-order optical field ($E^{2\omega}$) parallel and perpendicular to the pump polarization are $ E^{2\omega}_{\parallel} \propto -P_{x}\text{cos}(\theta) + P_{z}\text{sin}(\theta)$ and $ E^{2\omega}_{\perp} \propto P_{y}$, respectively. 
The corresponding SHG intensity $I^{2\omega} \propto (E^{2\omega})^2$ is characterized by an effective second-order susceptibility $\chi_\text{eff}^{(2)}$ that quantifies the symmetry breaking strength together with tensor components $d_{il}$. 
We measure $\chi_\text{eff}^{(2)}$ by fabricating Au nanodimers on bilayer MoS$_2$ supported on a 200-nm Au substrate (Fig.~\ref{Figure1}(c)) and characterize the  plasmonic structures via their dark-field scattering spectra (Fig.~S2). 
While pristine bilayer MoS$_2$ ($D_\text{3d}$ symmetry) exhibits negligible SHG intensity (Fig.~\ref{Figure2}(a), black line), depositing Au nanodimers on MoS$_2$ breaks this centrosymmetry through interfacial charge transfer, even without external excitation. 
This interfacial charge redistribution creates a local electric field at the Au-MoS$_2$ interface, leading to SHG signals. 
Under 532-nm CW excitation (resonant with the Au nanodimer), where the plasmonic resonance enables strong field enhancement and efficient carrier generation, the SHG intensity increases with pump power but saturates above 100 nW. 
The extracted $\chi^{(2)}_\text{eff}$ exhibits a nonlinear dependence on CW pump power (Fig.~\ref{Figure2}(b)). 
This saturation stems from the  interfacial charge accumulation that creates an energy barrier that limits further hot-electron injection.

We propose a capacitor model, based on our observations, where the electrons that accumulate at the metal/TMDCs interface create a dynamic energy barrier affecting subsequent hot-electron injection. 
We define $Q_{\rm inj}$ as the number of injected electrons, $Q_{\rm tot}$ as the total number of generated hot electrons, $\phi_b$ is the interfacial energy barrier, and the injection efficiency is $\eta=Q_{\rm inj}/Q_{\rm tot}$. 
The interface capacitance produces an energy barrier $\phi_b \propto Q_{\rm inj}$. 
In accordance with the Fowler rule for interfacial electron injection \cite{fowler1931analysis, PhysRev.127.2006, blandre2018limit}, we obtain $\eta \propto (\hbar\omega-cQ_\text{inj})^2$, where $c$ characterizes the capacitive barrier. 
By assuming that the incident power density is proportional to $Q_{\rm tot}$ and $\chi^{(2)}_\text{eff}$ is proportional to $Q_{\rm inj}$ (see details in Section 3, SM), we establish a direct relationship between $\chi^{(2)}_\text{eff}$ and the incident power density that agrees well with our experimental data (Fig.~\ref{Figure2}(b)). 
The energy barrier increases with $Q_{\rm inj}$ at higher laser power densities, explaining the reduced injection efficiency and consequent saturation of $\chi^{(2)}_\text{eff}$ (Fig.~\ref{Figure2}(c)). 
Additionally, we employ a drift-diffusion model to analyze the spatial distribution of injected hot electrons (Fig.~S5). 
While the time-resolved carrier spatial distribution depends on material-specific parameters that determine carrier decay rates, the distribution  shape remains independent of the number of injected electrons. 
This spatial consistency complements our capacitor model  by confirming that the charge distribution pattern remains consistent regardless of injection power, with only the magnitude changes. 
This validates our approach by demonstrating that interface charge accumulation can be modeled without considering complex power-dependent distribution shape variations.

Having demonstrated hot-electron induced symmetry breaking in bilayer MoS$_2$, we now examine the detailed characteristics of this symmetry transformation. 
Polarization-resolved SHG serves as a sensitive probe for analyzing the anisotropic modification of nonlinear coefficients induced by hot-electron injection, revealing distinct responses among tensor components and their relative strengths.
From Eq.~\ref{eq1}, the parallel component of the second-harmonic intensity can be expressed as:
\begin{equation}
    \begin{split}
        I_{\parallel}^{2\omega} \propto \left(\cos (\theta)\left( \sin (3 \phi)-\frac{d_{15}}{d_{22}} \sin (2 \theta)\right) \right. \\
        \left. +\sin (\theta)\left(\frac{d_{31}}{d_{22}} \cos (\theta)^2+\frac{d_{33}}{d_{22}} \sin (\theta)^2\right)\right)^2.
    \end{split}
    \label{eq2}
\end{equation}

Eq.~\ref{eq2} reveals that $I_{\parallel}^{2\omega}$ depends on the ratios between different $d_{ij}$ components ($d_{15}/d_{22}$, $d_{31}/d_{22}$, $d_{33}/d_{22}$) at specific angles $\theta$ and $\phi$.
In our experimental configuration, a high-NA objective (NA = 0.95) introduces significant z-direction field components (related to  $\theta$). 
Following established theory \cite{novotny2012principles} and  previous work \cite{li2021light}, focusing an $x$-polarized Gaussian beam generates field components of 0.905$E_0$ and 0.414$E_0$ in the $x$ and $z$ directions, respectively—equivalent to oblique incidence at $\theta \approx 24.6^\circ$. 
Fig.~\ref{Figure3}(a) displays SHG patterns for various ratios (0.2, 0.4, 0.6, 0.8) for this case; these results show the maximum intensity variations  occurs at specific azimuthal angles ($\phi = 30^\circ$, $150^\circ$, $270^\circ$ and $\phi = 90^\circ$, $210^\circ$, $330^\circ$). 
While incresing all three ratios contribute to the six-to-three petal pattern transformation, their impacts differ significantly: the $d_{15}$ component dominates, allowing us to focus primarily on its contribution while neglecting $d_{31}$ and $d_{33}$ in the subsequent analysis.

We incorporated a linear polarizer in the detection path to isolate SHG signals parallel to the incident femtosecond laser polarization ($I_{\parallel}^{2\omega}$) in the polarization-resolved SHG measurements. 
Rotating the sample varies the azimuthal angle $\phi$ between the incident polarization and crystal axis. 
Without CW excitation, the SHG pattern shows a characteristic two-lobed distribution (Fig. S6(a)) arising from anisotropic plasmonic resonances in the gold nanodimer \cite{wang2024symmetry}. 
The structural anisotropy of the gold nanodimer enhances plasmonic modes when the incident polarization aligns with its long axis, producing stronger SHG intensity parallel to this axis compared to perpendicular orientations. 
This plasmon-driven polarization difference dominates intrinsic MoS$_2$ symmetry features. 
By normalizing SHG intensities with (Fig. S6(b)) and without (Fig. S6(a)) CW excitation (Fig. 4(c)), we disentangle hot-electron-induced symmetry breaking from the plasmonic background. 
Fitting these data to our theoretical model yields $d_{15}/d_{22} \approx 0.2$; this quantifies the symmetry-breaking anisotropy strength. 
This ratio compares $d_{15}$ (an in-plane component with perpendicular field coupling) with $d_{22}$ (a purely in-plane component), directly measuring how hot-electron injection anisotropically modifies the nonlinear response.
Our measurements establish tensor component ratios as effective metrics for characterizing symmetry-breaking anisotropy strength in 2D materials.

Density functional theory calculations were performed to understand the mechanism of hot-electron induced symmetry breaking in bilayer MoS$_2$ using the \textit{VASP} code \cite{kresse1996efficient,perdew1996generalized,blochl1994projector}. 
Asymmetric charge distribution was modeled by applying an perpendicular electric field to a 2H-stacked bilayer MoS$_2$ unit cell following methodology established earlier \cite{PhysRevB.109.075417, guan2024giant}. 
Self-consistent calculations revealed that external electric fields induce asymmetric charge distributions between layers in which the interlayer charge difference scales linearly with field strength (based on our Bader charge analysis results, Fig.~S7). 
The linear relationship establishes a direct correspondence between applied field strength and hot-electron injection, providing a theoretical foundation for our experimental observations.

\begin{figure}[h!]
    \centering
    \hspace{-0.2cm}\includegraphics[scale=0.48]{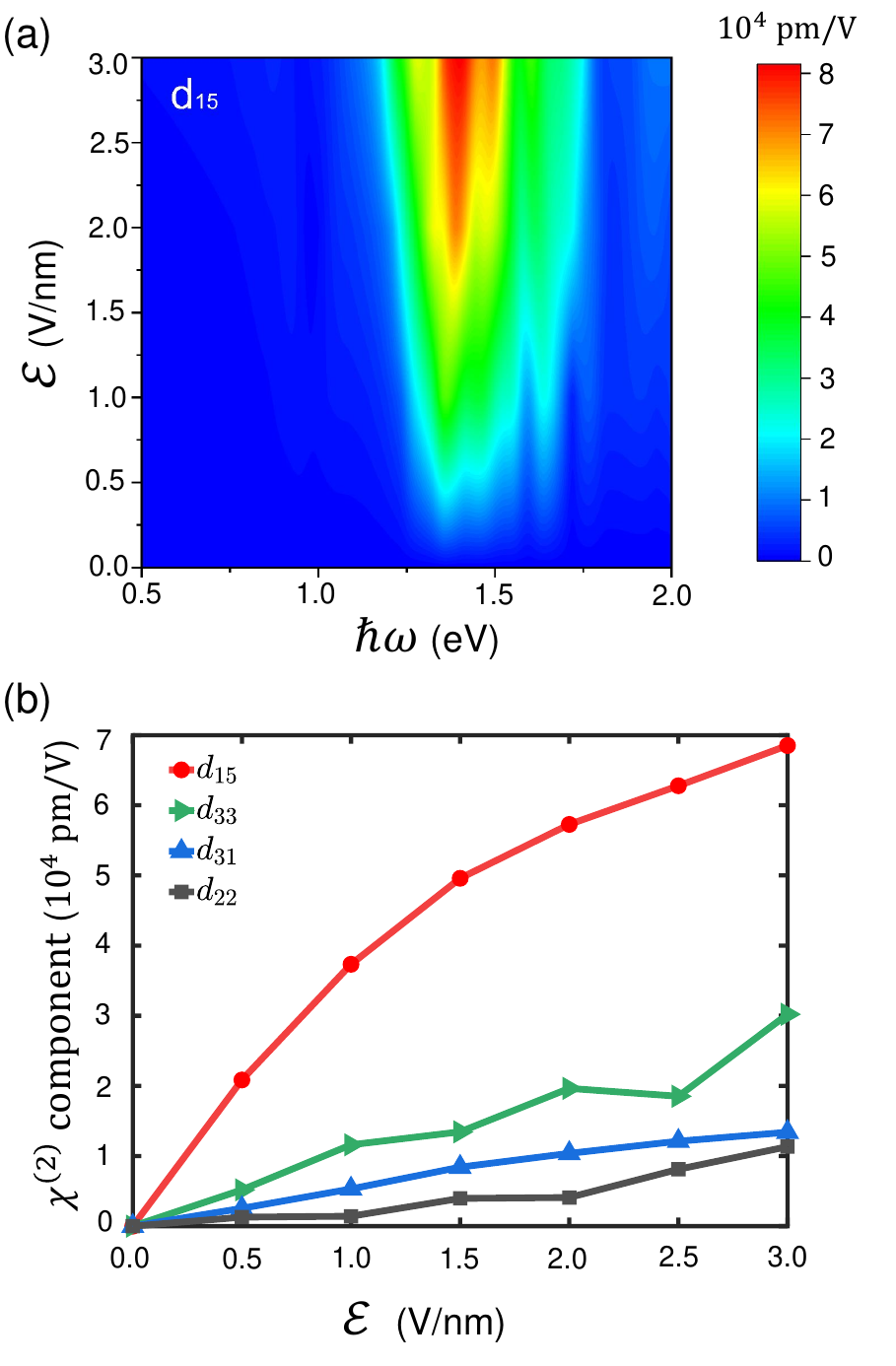}
    \caption{(a) DFT-calculated $d_{15}$ dependence on external electric field strength $\mathcal{E}$ and pump photon energy $\hbar\omega$. Other tensor components ($d_{22}$, $d_{31}$, $d_{33}$) are shown in Fig.~S4. (b) Electric-Field-dependent susceptibility tensor components under resonance conditions at $\hbar\omega \approx 1.45$ eV.}
    \label{Figure4}
\end{figure}

Analysis of nonlinear optical tensor components reveals that all components exhibit strong field and photon energy dependence around the incident wavelength, with significant resonance enhancement at $\sim$ 1.45 eV (Fig.~4(a), Fig.~S4) \cite{guan2024giant}. At this resonance, where twice the incident energy approaches the C-exciton energy \cite{lee2015resonant}, all tensor components vanish at zero field but respond differently as field strength increases. 
The $d_{15}$ component increases most rapidly, confirming our experimental finding that its ratio with $d_{22}$ dominates the SHG response. However, our calculations predict that$d_{15}>d_{22}$, contrary to our experimental measurements (i.e., $d_{15}$ is only $\sim0.2d_{22}$).
This discrepancy stems from limitations in our DFT model, in which a uniform electric field was applied to a bilayer MoS$_2$ unit-cell such that lateral carrier diffusion effects do not occur. In real samples, hot carriers accumulate at the interface and diffuse both vertically and laterally, potentially inducing laterally inhomogeneous electric fields, reducing the symmetry to below C$_{3v}$ and further enhancing $d_{22}$. Despite these limitations, our theoretical analysis and DFT calculations provide robust support for the observed symmetry transformation, while indicating that larger simulation models and time-dependent approaches are needed to accurately describe the competition between different susceptibility components to better understand the symmetry breaking mechanism. 

\textit{Conclusion.}---This work demonstrates symmetry breaking in bilayer MoS$_2$ via hot-electron injection. 
Theoretical analysis reveals that the asymmetric charge distribution induced by hot electrons creates effects analogous to those from an applied electric field. 
Plasmonic hot-electron injection was achieved in a configuration consisting of a Au-nanodimer on bilayer-MoS$_2$ atop a Au-film substrate. 
Power-dependent SHG measurements revealed saturation behavior at higher powers. 
This behavior is well-described by a capacitor model in which interfacial charge accumulation creates a dynamic energy barrier that limits further electron injection. 
Polarization-resolved SHG measurements quantify the anisotropic modification of nonlinear susceptibility, with component ratio $d_{15}/d_{22} \approx 0.2$ serving as an effective metric for symmetry-breaking anisotropy. First-principles calculations show that perpendicular electric fields preferentially enhance the $d_{15}$ component, confirming experimental observations while highlighting the need for larger-scale simulation models and time-dependent calculations to better describe the symmetry-breaking mechanism.
Our work establishes SHG as an effective probe of hot-electron-induced symmetry breaking, providing insights into ultrafast charge-mediated symmetry controlling. 
These methods also enable future investigations of non-local, ultrafast electron dynamics in different metal-semiconductor plasmonic heterostructures. 

\vspace{1ex}

\textit{Acknowledgements.}---We gratefully acknowledge financial support by the RGC of Hong Kong (ZP and DL $-$ GRF Grant No. 15304519, DJS and GZ $-$ GRF Grant No. 1122534). 
ZG also acknowledges the computational resource provided by Tianhe Xingyi from National Supercomputer Center in Guangzhou.

\bibliography{reference.bib}

\end{document}